\documentclass[aps,prd,twocolumn,showpacs,superscriptaddress,groupedaddress,nofootinbib]{revtex4}   
\usepackage[utf8]{inputenc}
\usepackage{amsmath}
\usepackage{amsfonts}
\usepackage{graphicx}
\usepackage{multirow}
\usepackage{xcolor}
\usepackage{hyperref}
\usepackage{float}
\usepackage{ulem}

\begin{document}
\title{Lessons from HAWC PWNe observations: the diffusion constant is not a constant; Pulsars remain the likeliest sources of the anomalous positron fraction; Cosmic rays are trapped for long periods of time in pockets of inefficient diffusion
}

\author{Stefano Profumo}
\email{profumo@ucsc.edu}
\affiliation{Department of Physics, University of California, 1156 High St. Santa Cruz, CA 95060, United States of America}
\affiliation{Santa Cruz Institute  for Particle Physics, 1156 High St. Santa Cruz, CA 95060, United States of America
}
\author{Javier Reynoso-Cordova}
\email{reynosoj@fisica.ugto.mx}
\affiliation{Santa Cruz Institute  for Particle Physics, 1156 High St. Santa Cruz, CA 95060, United States of America
}
\affiliation{
Departamento de F\'isica, DCI, Campus Le\'on, Universidad de
Guanajuato, 37150, Le\'on, Guanajuato, M\'exico}
\author{Nicholas Kaaz}
\email{nomahen@ucsc.edu}
\affiliation{Department of Physics, University of California, 1156 High St. Santa Cruz, CA 95060, United States of America}
\author{Maya Silverman}
\email{massilve@ucsc.edu}
\affiliation{Department of Physics, University of California, 1156 High St. Santa Cruz, CA 95060, United States of America}

\begin{abstract}
Recent TeV observations of nearby pulsars with the HAWC telescope have been interpreted as evidence that diffusion of high-energy electrons and positrons within pulsar wind nebulae is highly inefficient compared to the rest of the interstellar medium. If the diffusion coefficient well outside the nebula is close to the value inferred for the region inside the nebula, high-energy electrons and positrons produced by the two observed pulsars could not contribute significantly to the local measured cosmic-ray flux. The HAWC collaboration thus concluded that, under the assumption of isotropic and homogeneous diffusion, the two  pulsars are ruled out as sources of the anomalous high-energy positron flux. Here, we argue that since the diffusion coefficient is likely {\it not} spatially homogeneous, the assumption leading to such conclusion is flawed. We solve the diffusion equation with a radially dependent diffusion coefficient, and show that the pulsars observed by HAWC produce potentially perfect matches to the observed high-energy positron fluxes. We also study the implications of inefficient diffusion within pulsar wind nebulae on Galactic scales, and show that cosmic rays are likely to have very long residence times in regions of inefficient diffusion. We describe how this prediction can be tested with  studies of the diffuse Galactic emission.
 \end{abstract}

\maketitle

\section{Introduction}\label{sec:Introduction}

The origin of the cosmic radiation and the mechanisms responsible for cosmic-ray acceleration remain largely, and in detail, unknown. Over the last few years, the discovery by the PAMELA (Payload for Antimatter Matter Exploration and Light-nuclei Astrophysics) detector that the positron fraction, the ratio of the positron to positron-plus-electron flux in the cosmic radiation, {\it increases} with energy for particle energies greater than about 10 GeV is especially puzzling \cite{Adriani:2010rc}. This result was subsequently confirmed by the Fermi LAT (Large Area Telescope) \cite{Ackermann:2010ij} and by AMS (the Alpha Magnetic Spectrometer) \cite{PhysRevLett.110.141102}, with increasingly large and convincing statistics.

An increasing high-energy positron fraction is puzzling because in almost any predictive Galactic cosmic-ray model (see e.g. \cite{Strong:1998pw}), secondary-to-primary ratios, such as the positron fraction, or B/C, must asymptotically decline with energy --  simply as a result of the energy-dependence of the diffusion coefficient, which, in turn, is a macroscopic manifestation of the energy dependence of the Larmor radius. A growing positron fraction is thus likely a signal that the observed ``anomalous'' positrons, whose energy is observed up to almost 1 TeV, are not of secondary origin but, rather, are produced as primary particles.

High-energy positrons and electrons cool very efficiently, both via synchrotron and via inverse Compton, both processes responsible for a rate of energy loss quadratic in the energy of the particle. The cooling time of a TeV positron in the Galaxy is, typically, on the order of $$\tau_e\sim 3\times 10^5\ {\rm yr}\times (1\ {\rm TeV}/E_e).$$ Cosmic-ray electrons and positrons diffuse in a random-walk pattern in the magnetic field of the Galaxy. If this process is described with a uniform diffusion coefficient $D$, sources producing TeV positrons observed at Earth must lie within a distance $$d\lesssim \sqrt{D \tau_e}\sim 0.5\ {\rm kpc}$$ for a standard diffusion coefficient \cite{Strong:1998pw} $$D_{\rm ISM}\sim 3\times 10^{28}\ {\rm cm}^2/{\rm s}\ (E_e/{\rm GeV})^{0.33}.$$ High-energy positrons (and electrons) therefore can only be produced within a few kpc (at most) sphere of our Galactic position.

Of the potential sources of high-energy cosmic-ray positrons and electrons, one stands out as especially intriguing: the annihilation of Galactic dark matter (see e.g. \cite{Cholis:2013psa,Bergstrom:2008gr,Cirelli:2008jk,Nelson:2008hj,ArkaniHamed:2008qn,Cholis:2008qq,Harnik:2008uu,Fox:2008kb,Pospelov:2008jd,MarchRussell:2008tu,Chang:2011xn}): In many microscopic realizations, dark matter particles can pair annihilate into matter-antimatter pairs, such as positrons and electrons. The annihilation rate, additionally, can be in some cases associated with the observed abundance of dark matter. This possibility has received considerable interest ever since the PAMELA result, but is very tightly constrained by the fact that no associated signal has been observed in gamma rays from dark matter annihilation, for example, in local dwarf spheroidal galaxies \cite{Ackermann:2011wa} or in nearby clusters of galaxies \cite{Ackermann:2010qj} 

For many years, an alternate, more mundane explanation has also been put forward \cite{Hooper:2008kg,Yuksel:2008rf,Profumo:2008ms,Malyshev:2009tw,Grasso:2009ma,Linden:2013mqa,Cholis:2013psa,Linden:2013mqa}: that middle-aged ($\tau\sim 10^5-10^6$ yr), nearby pulsars could accelerate, in their magnetosphere and, subsequently, in the surrounding shock with the interstellar medium, (primary) electrons and positrons to very high energy. The energetics of known candidate pulsars, as indicated by the pulsars' spin-down luminosity, is in the correct range to explain the observed excess positrons, as long as (1) the diffusion coefficient is $D\sim D_{\rm ISM}$ between the pulsar and Earth, and (2) a fraction of ${\cal O}(10\%)$ of the spin-down luminosity is injected in positron-electron pairs \cite{Profumo:2008ms,Linden:2013mqa}.

Recently, the HAWC (High-Altitude Water Cherenkov) Observatory confirmed earlier results from Milagro \cite{Abdo:2009ku} and HAWC \cite{Abeysekara:2017hyn} and observed extended TeV emission surrounding two nearby pulsars, Geminga and Monogem (PSR B0656+14), among the candidate sources for the observed anomalous high-energy positrons \cite{Abeysekara:2017old}. As previously noted in \cite{Hooper:2017gtd}, the spectrum and morphology from TeV gamma-ray observations can be used to infer features of the underlying high-energy electron-positron population responsible for the up-scattering of photons to TeV energies. In fact, Ref.~\cite{Hooper:2017gtd} argued that HAWC observations available at the time \cite{Abeysekara:2017hyn} were entirely compatible, and actually supported, the hypothesis that the observed excess high-energy positrons originated from nearby pulsars.

The recent HAWC observations \cite{Abeysekara:2017old} provide detailed information on the spatially extended emission of high-energy, TeV gamma rays from the pulsar wind nebula (PWN) regions surrounding Geminga and Monogem. Since hadronic processes are very unlikely given the environmental gas density in PWNe \cite{1984AJ.....89.1022P}, the emission traces the brightness of inverse Compton emission from high-energy electrons and positrons.  As a result of the large Klein-Nishina suppression for the electron-positron scattering off of high-energy photons, the target photon population is predominantly given by the cosmic microwave background, with a spatially uniform density distribution \cite{Abeysekara:2017old}. As a result, the brightness profile of the TeV emission potentially provides direct information on the radial distribution of the high-energy electron-positron population in the observed PWNe.

Following \cite{Abeysekara:2017old}, we have also fit the brightness profile of the TeV emission, and indeed confirmed that (1) the emission is well fit by a purely diffusive density distribution for the electron and positron population, (2) the best-fit outward wind value is zero, meaning there is no evidence for non-diffusive transport outwards, and (3) the preferred inferred value of the diffusion constant at the energies of interest is significantly smaller than what inferred from standard cosmic-ray propagation models. We also concur with the HAWC analysis that the diffusion coefficient needed to fit the data is significantly lower than what inferred from hadronic cosmic rays in the Galaxy \cite{Strong:1998pw}.

A Bayesian search for best-fit parameters applied to the GALPROP package \cite{Strong:1998pw} indicated that different cosmic ray species likely probe very different regions of the interstellar medium, implying that the standard strategy of relying on e.g. the Boron-to-Carbon (B/C) ratio to calibrate propagation parameters might be unreliable \cite{Johannesson:2016rlh}. This analysis, however, does not indicate that a difference in the propagation parameters between leptonic and hadronic cosmic rays should exist. Rather, it indicates that homogeneity is likely an incorrect assumption when large-scale Galactic diffusion is considered. Additional support for the lack of homogeneity in the diffusive properties of the interstellar medium comes from magnetohydrodynamic simulations \cite{Yan:2004aq} and from models of cosmic ray escape from supernova remnants \cite{Malkov:2012qd}, which generically indicate an exponential suppression of the diffusion coefficient inside the expanding cloud around cosmic-ray accelerators.

The HAWC Collaboration \cite{Abeysekara:2017old} entertained the possibility that the diffusion coefficient inferred from TeV observations is constant (homogeneous and isotropic) all the way to Earth. If that is the case, the contribution to the local flux of cosmic-ray positrons and electrons from Geminga and Monogem, the two pulsars whose nebulae were observed in their TeV emission, is highly suppressed and essentially negligible \cite{Abeysekara:2017old}. Ref.~\cite{Abeysekara:2017old} shows that such conclusion holds for a broad range of choices for the pulsar initial spin-down timescale, the energy dependence of the diffusion coefficient as well as the spectral index and flux normalization of the injected cosmic rays.

Hooper and Linden \cite{Hooper:2017tkg} argued that the assumption of isotropic and homogeneous diffusion inside and outside the pulsar nebulae is inconsistent with the detection of high-energy (up to 20 TeV) electrons with H.E.S.S. \cite{Abdalla:2017brm}. The key argument is that a 20 TeV electron cools in around $t_{\rm cool}\sim 10,000$ years, and in that time such particles can only travel on the order of $\sqrt{D\cdot t_{\rm cool}}$ which, for the diffusion coefficient inferred by the HAWC observations inside the nebulae would indicate the presence of a high-energy cosmic-ray source within 10-20 pc of the Sun's position. Plausible candidates for such a powerful, nearby source, however, are not known to exist: if such a source existed, it should in fact be readily detectable from its high-energy photon emission \cite{Hooper:2017tkg}.

Interestingly, previous theoretical studies have suggested that cosmic-ray gradients can induce suppressed diffusion inside supernova remnants \cite{Malkov:2012qd}. Alfv\'en waves generated by cosmic rays induce a net force that suppresses diffusion near the sites of cosmic-ray acceleration and, more generally, where cosmic-ray fluxes are larger \cite{1981A&A....98..195A}. Subsequent studies have elaborated on the role of non-linear diffusion of cosmic rays generated by magnetohydrodynamic turbulence around acceleration sites. Such turbulence induces an effective diffusion coefficient which depends on the cosmic-ray density and is far from spatially homogeneous \cite{Ptuskin:2008zz}. Ref.~\cite{DAngelo:2015cfw} studies the effect of cosmic-ray induced inhomogeneity in the diffusion coefficient from the gradient in cosmic-ray density induced by the large particle densities in the disc (``near-source regions'') versus the much lower densities in the diffusive halo at large Galactic latitudes; They conclude that the residence time of cosmic rays in regions of inefficient diffusion in the thick Galactic disk is much larger than simple estimates based on global grammage such as the boron-to-carbon ratio. Ref.~\cite{Nava:2016szf}, following the earlier work of \cite{Malkov:2012qd}, studies in a self-consistent picture the back-reaction on transport properties of cosmic-ray out-flux from supernova remnants, showing numerically that streaming instabilities significantly suppress the diffusion coefficient in regions tens of pc from the acceleration site, and for timescales, for high-energy cosmic rays of several TeV of energy, up to millions of years. Finally, ref.~\cite{DAngelo:2017rou} studies observational consequences of the ``self-confinement'' of cosmic rays near their acceleration sites on the diffuse gamma-ray background, using Galactic population models for supernova remnants. Their conclusion are close to what Hooper and Linden find: the fraction of volume of the diffusive halo occupied by pockets of inefficient diffusion is small (less than 1\%) if the radius of such pockets is on the order of 30 pc, although it could be substantial if the pockets of inefficient diffusion are on the order of 150 pc \cite{Hooper:2017tkg}.

In this study, we build on previous theoretical work, and utilize the HAWC obseravtions to study the effect of non-homogeneous, non-isotropic diffusion near cosmic-ray acceleration sites, and to re-assess whether the HAWC result affects the plausibility of local pulsars as the culprit for the observed positron excess. Our key findings are that physically plausible and theoretically motivated models for the radial dependence of the diffusion coefficient inside and outside the pulsar wind nebulae of Geminga and Monogem likely lead to significant fluxes of high-energy electrons and positrons from those sources to Earth. Models of the Galactic pulsar population and observationally-backed assumptions on the size of pulsar wind nebulae with time indicate that Galactic cosmic rays spend a very significant fraction of their residence time in inefficient diffusion regions; we discuss and propose several observational tests that will allow to test this hypothesis.  

The remainder of this work is structured as follows: in Section \ref{sec:Solution} we provide details on the algorithm we use to solve the 3D diffusion equation and we describe the parameters and structure of the diffusion models we employ; in Section \ref{sec:results} we present our results on the differential positron flux; in Section \ref{sec:Macroscopic_effects} we discuss the implications of pockets of inefficient diffusion on large scales and ways to probe their existence, and, finally, we summarize and discuss our results in Section \ref{sec:Discussion}.

\section{Numerical solution of cosmic-ray diffusion with a non-homogeneous diffusion coefficient}\label{sec:Solution}

We describe cosmic-ray diffusion in the Galaxy, as customary, through the partial differential equation 
\begin{equation}
\frac{\partial \psi}{\partial t} = \vec{\nabla}\cdot(D(\vec{x},E) \vec{\nabla} \psi) + \frac{\partial}{\partial E}(P(E) \psi) + Q,
\label{eq:diffeq}
\end{equation}
where the unknown function $\psi$ indicates the differential number of particles in energy and volume (i.e. the differential number density in energy), and is a function of time, position, and energy, $$\psi = \psi(t,x,E)=\frac{d^4N}{dEdV},$$
$D(\vec{x},E)$ is the diffusion coefficient, which we take to be time-independent but spatially inhomogeneous, $P(E)$ describes energy losses (which we assume homogeneous and isotropic), and, finally, $Q$ is the injection source term. In this work we are interested, for simplicity, in point-like injection sources in time and space, i.e. $$Q \propto \delta(\vec{x}-\vec{x}_o)\delta(t - t_o).$$ 

We assume spherical symmetry for the spatial dependence of the diffusion coefficient, $$D(\vec{x},E)=D(r,E),$$ and solve the differential equation in radial shells with a width such that in each shell $i$ at radius $r_i$ the diffusion coefficient can be considered approximately constant\footnote{We elaborate below on the case of a sudden step-like change in the diffusion coefficient, which we also consider.}, $D(r_i,E)=D_i(E)$. In each shell, the differential equation (\ref{eq:diffeq}) has a known Green's function \cite{PhysRevD.52.3265}, which can be derived as follows: 

(i) change $\psi \to rPf$; 

(ii)  define $T = \int_{E}^{E_o} 1/P(x) dx$, i.e. the time it takes for a particle of initial energy $E_o(E)$ to cool down to an energy $E$; 

(iii) define $\tau = T - t$, $z = T$; and, finally, 

(iv) introduce the variable $$u_i = \int_0^z D_i(x) dx.$$ 
In each spherical shell, steps (i)--(iv) yield a standard 
one-dimensional diffusion equation with unit diffusion coefficient,
\begin{equation}\label{eq:modeq}
\frac{\partial f}{\partial u} = \frac{\partial^2 f}{\partial r^2}.
\end{equation}
For spatially constant diffusion coefficients, Eq.~(\ref{eq:modeq}) has the following  Green's function, switching back to $\psi$, valid for an arbitrary injection spectrum $N_o$,
\begin{equation}
\psi(t,r,E)=\frac{N_o(E_o) P(E_o)}{\pi^{3/2} P(E) r_{\rm{diff}}^3 }e^{-r^2/r_{\rm{diff}}^2},
\label{eq:analytical_solution}
\end{equation}
where 
\begin{equation}
r_{\rm{diff}}^2 =4\int_{E}^{E_o} D(x)/P(x) dx=4\Delta u,
\label{eq:rdiff}
\end{equation}
and where $N_o(E_o)$ is the injection spectrum, evaluated at the energy $E_o$. The initial injection spectrum here is assumed to be parametrized by a power law $Q \propto E^{-\alpha}$, possibly with a high-energy exponential cutoff. The solution for a generic initial condition is a convolution integral of Eq.~(\ref{eq:analytical_solution}) with the given initial condition. Alternately, the equation can also be solved using Monte Carlo methods \cite{ikeda}. These methods have the significant advantage of easily and efficiently generalizing to the case of non spatially homogeneous diffusion coefficients. We now describe the numerical procedure we adopt, and provide details on the numerical accuracies compared to the Green's function method in the Appendix.

Our Monte Carlo integration routine is as follows: First, we inject particles at the pulsar's location, for convenience set at the origin of the coordinate system, with initial energy $E_o(E)$, which we compute from the integration of a purely quadratic energy loss form for $P(E)=dE/dt$ (we assume that the relevant processes are inverse Compton scattering and synchrotron with $P(E)\simeq P_o E^2$, with $P_o \sim 1.02 \times 10^{-16}\ {\rm GeV}/{\rm s}$; notice that we also implemented Klein-Nishina corrections to the inverse Compton energy losses; we found very small effects, which came at great computational cost; we therefore neglected such corrections for most of the results shown below). {The resulting initial energy for an electron produced a time $t_0$ ago is then
\begin{equation}
E_o (E) = \frac{E}{1- P_o E t_o},\ \ E\le E_{\rm max}=\frac{1}{P_ot_o},
\end{equation}
where $E_{\rm max}$ indicates the maximal energy attainable for an electron produced a time $t_0$ ago, and with $E_o(E>E_{\rm max})$ not defined (effectively cutting off the range for $E_o$).}
We assume that the energy dependence of the diffusion coefficient reads
\begin{equation}\label{eq:dzero}
D(E)=D_o \left(\frac{E}{\rm{GeV}} \right)^\delta,
\end{equation}
with typical values of $D_o \sim 10^{28}\ \rm{cm}^2 \rm{s}^{-1}$ and $\delta = 1/3$ as determined e.g. in Ref.~\cite{PhysRevD.52.3265,Hooper:2017tkg,Feng:2015uta}.

{Notice that for the specified functional forms for the diffusion and energy loss functions, it is straightforward to analytically compute, for a given injection time $t_o$, the corresponding $r_{\rm diff}(E)$ for $E<E_{\rm max}=1/(P_o\ t_o)$, which reads
\begin{equation}\label{eq:rdiffana}
r^2_{\rm diff}(E)=\frac{4D_o}{P_o}\frac{E^{\delta-1}}{1-\delta}\left(1-\frac{1}{\left(1-P_oEt_o\right)^{\delta-1}}\right).
\end{equation}

The maximal diffusion radius corresponds to the maximal energy for a burst-like injection at time $t_o$, i.e. $E_{\rm max}=1/(P_o t_o)$, and reads
\begin{equation}\label{eq:rdiffmax}
r^2_{\rm diff,\ max}=\frac{4D_o}{P_o}\frac{\left(P_o t_o\right)^{1-\delta}}{1-\delta}.
\end{equation}
}

For a given initial and final energy, the spatial evolution of the particle is simulated via a Brownian random walk of diffusion radius $r_{\rm diff}$,
where the diffusion radius is calculated using the diffusion coefficient corresponding to the particle's initial location \cite{rogers}. The particle position is then evolved in each spatial dimension $(\vec x)_k$, $k=1,2,3$  in Cartesian space according to the prescription 
\begin{equation}
(\vec {x^\prime})_k=(\vec x)_k+\eta\left(2\sqrt{\Delta u_i}\right),
\end{equation}
where the subscript $i$ labels the spherical shell $i$ with constant diffusion coefficient $D_i$ corresponding to the particle's initial position $\vec x$, and where $\eta$ is a random number generated from a normal distribution with mean $\mu=0$ and width $\sigma=1$. The procedure is then repeated $N_{\rm steps}\gtrsim10^3$ times, the precise value depending on the initial and final energy, and the distance from the origin of the particle's final position, $|\vec x_f|$ is stored in radial bins. For each energy, the algorithm is repeated for $N_{\rm particles}\gtrsim 10^4$ particles until numerical convergence for the radial density profile is achieved. 

The validity of the use of a Gaussian kernel (corresponding to the Green's function of the spatially homogeneous diffusion coefficient case) is a general result of It\^o calculus  for the general Fokker-Planck advection-diffusion equation with non-constant coefficients \cite{ikeda, rogers}; the validity of evolving the system via a Gaussian kernel holds for sufficiently small ``time'' steps in the numerical evolution of the diffusing particles. Therefore, the validity of our approach depends on the ratio of the length scale associated with homogeneous diffusion over a time-step $\Delta t$, i.e. $$\Delta_{\rm hom}\sim \sqrt{D\Delta t},$$ and the length scale associated to the spatial variation of the diffusion coefficient in the radial direction, $$\Delta_{\rm non-hom}\sim |\partial D/\partial r|\Delta t.$$ The ratio $\Delta_{\rm non-hom}/\Delta_{\rm hom}\ll1$ corresponds to the requirement of the time step being $\Delta t\ll D/|\partial D/\partial r|^2$. We set $N_{\rm steps}$ above such that this criterion applies, with the exception of a step-like jump in $D(r)$. In the latter case, we simply evolve the particles according to the relevant diffusion coefficient within and beyond the discontinuity, where the solution to the diffusion equation is exact, and rely, at the boundary, on the fact that the Monte Carlo diffusion algorithm automatically ensures particle number and flux conservation, preventing artificial spurious production or destruction of the diffusing particles.

As explained in the Introduction, the recent HAWC observations indicate inefficient diffusion in the two observed pulsar wind nebulae, with a preferred value for the diffusion coefficient inside the Geminga PWN at 100 TeV of $3.2^{+1.4}_{-1.0}\times 10^{27}\ {\rm cm}^2{\rm s}^{-1}$ and inside the Monogem  (PSR B0656+14) PWN of $15^{+49}_{-9}\times 10^{27}\ {\rm cm}^2{\rm s}^{-1}$ \cite{Abeysekara:2017old}, to be contrasted with a putative diffusion coefficient for the interstellar medium (ISM) at large from GALPROP\footnote{http://galprop.stanford.edu/} analyses \cite{Strong:1998pw}, at the same energies, of about $10^{30}\ {\rm cm}^2{\rm s}^{-1}$; HAWC therefore indicates a suppression of the diffusion coefficient of a factor between 200-500 (Geminga) and (15-150) inside the PWN compared to the ISM large-scale value.

\begin{figure}[t]
\centering
\includegraphics[width=0.8\linewidth]{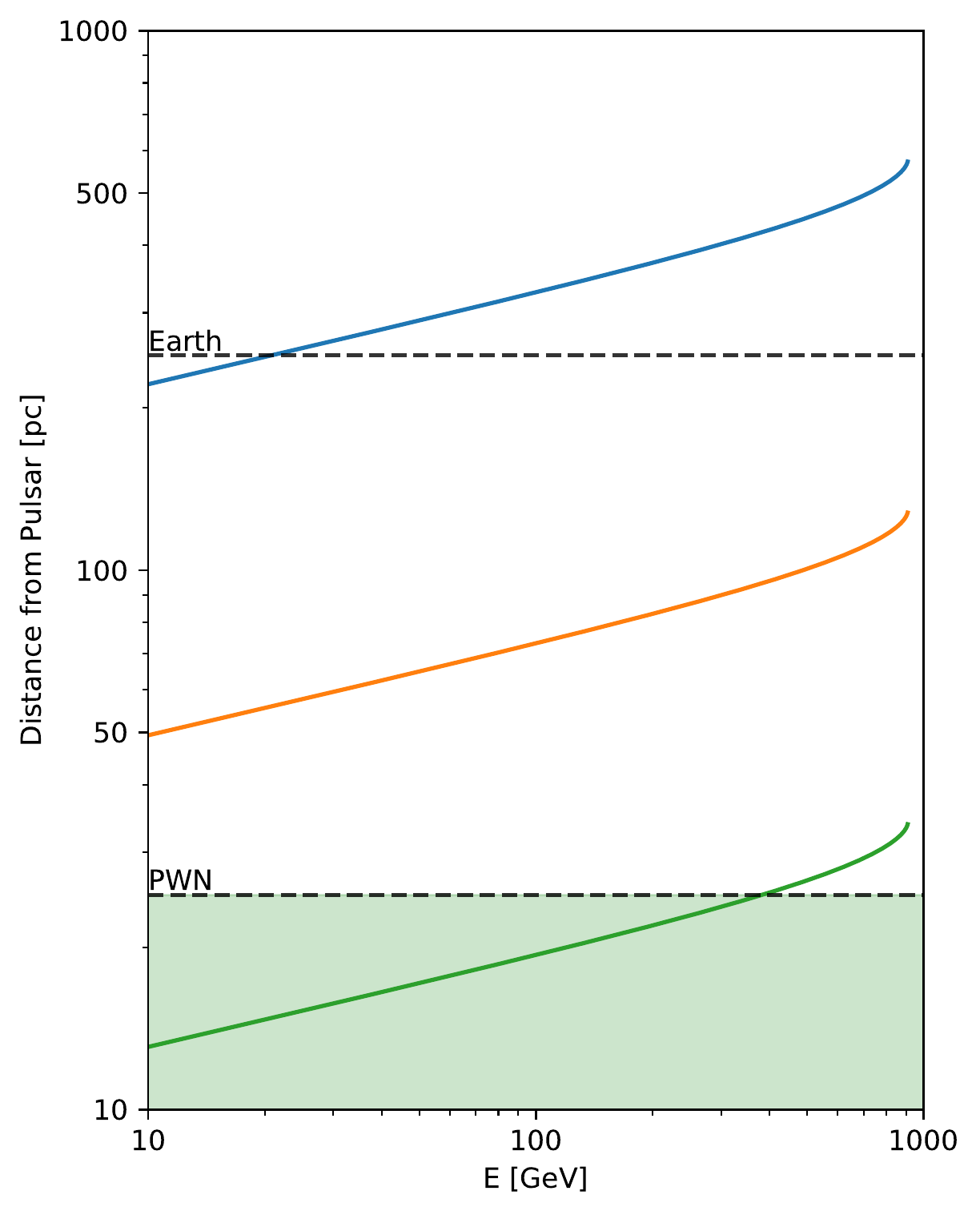}
\caption{{The diffusion length $r_{\rm diff}(E)$, for $E<E_{\rm max}=1/(P_o\ t_o)$ (see Eq.~(\ref{eq:rdiff}) and (\ref{eq:rdiffana})) for electrons/positrons, for different choices for the diffusion constant, for a pulsar of age $t_o=3.42 \times 10^{5} $ yr (the characteristic age of Geminga). A larger diffusion constant allows for larger distances, while a low diffusion constant barely allows any of the particles to escape the PWN region (shaded green).}} 
\label{Fig:diff_lenght}
\end{figure}

{Such inefficient diffusion yields strongly suppressed diffusion lengths, which following the Green's function solution can be defined as in Eq.~(\ref{eq:rdiff}). We plot in Fig.~\ref{Fig:diff_lenght} the diffusion length $r_{\rm diff}(E)$ as a function of energy for $E<E_{\rm max}=1/(P_o\ t_o)$, for three choices of the normalization of the diffusion coefficient $D_o$ (defined as in Eq.(\ref{eq:dzero}) and thus here corresponding to the value at 1 GeV of energy) of $D_o=200,\ 10,\ 0.7\times10^{26}\ {\rm cm}^2{\rm s}^{-1}$, for a burst-like injection occurring at a time $t_o=3.42 \times 10^{5} $ yr (the characteristic age of Geminga) in the past. The vertical axis indicates the distance from the pulsar. The curves on the plot are readily understood from Eq.~(\ref{eq:rdiffana}). The maximal diffusion length, independent of $D_o$, corresponds to roughly TeV energies for a pulsar of Geminga's age. This is in accordance with the analytical expression given in Eq.~(\ref{eq:rdiffmax}) above.}

Our figure illustrates that for values of the diffusion coefficient close to what indicated by the HAWC observations, $D_o\sim0.7\times10^{26}\ {\rm cm}^2{\rm s}^{-1}$, the diffusion length of the injected cosmic-ray electrons and positrons is barely larger than the physical size of the PWN (we shade here only to guide the eye the region at distances smaller than 20 pc as indicative of those distances likely still within the PWN). For intermediate diffusion coefficients the diffusion length is shorter than the distance separation to the Earth (but large enough that a non-trivial flux at Earth of cosmic rays injected from the pulsar is expected) and, finally, for values of $D_o$ close to those inferred by global cosmic-ray data the Earth is well within the diffusion length for a large interval of energies.

The outcome of the HAWC collaboration hypothesis that the diffusion coefficient inside and outside the PWN is constant, is therefore entirely unsurprising. However, as explained above, such hypothesis is also unrealistic and theoretically unmotivated. Here, we thus model different diffusive mediums with a simple model where within a few parsecs from the PWN particles experience inefficient diffusion, at the levels observed indirectly by HAWC, followed by a transition from the low diffusion to a high diffusion regime, with a diffusion constant close to the global ISM values.  

\begin{figure}[t]
\centering
\includegraphics[width=1\linewidth]{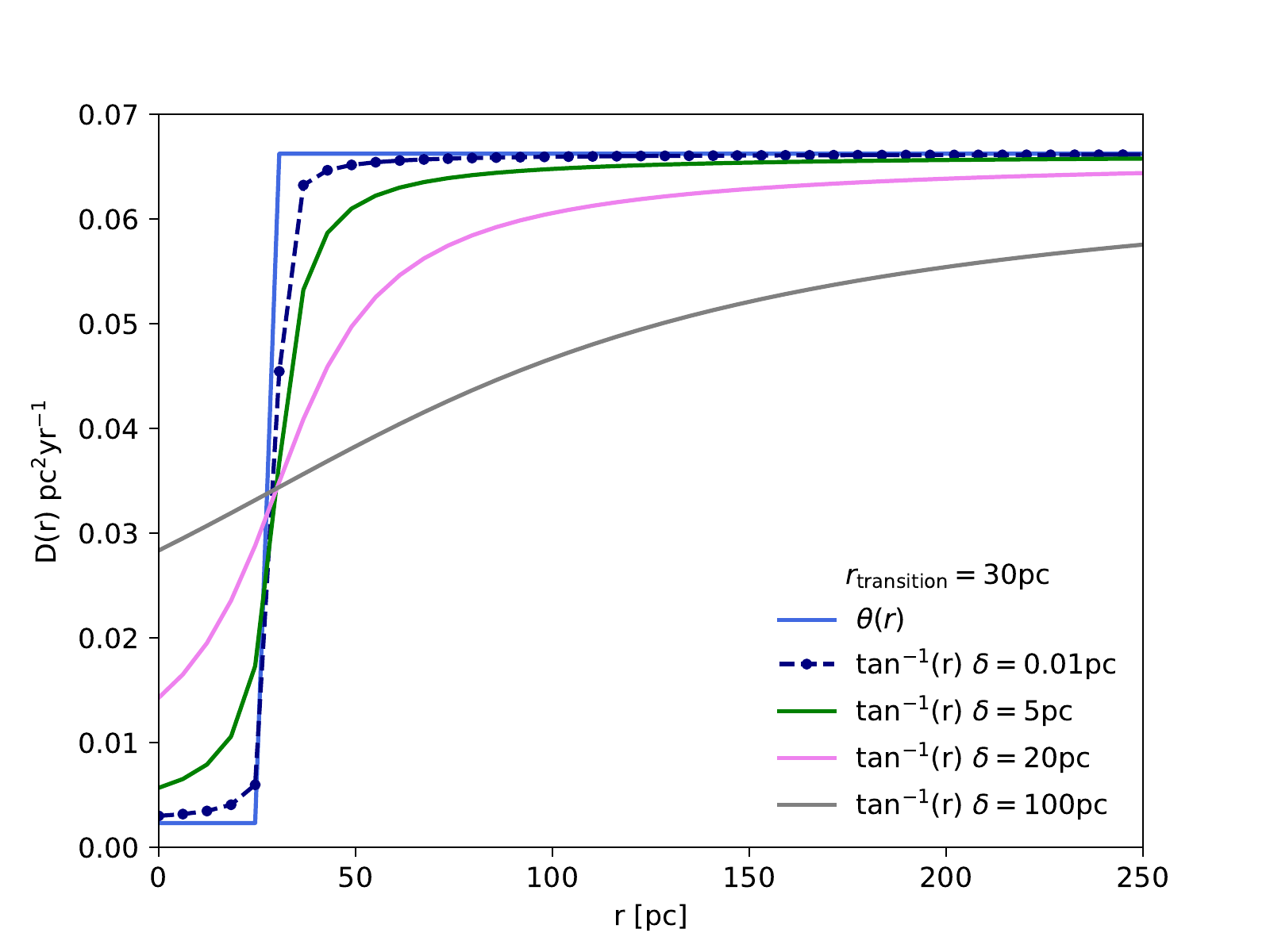}
\caption{The radial dependence of the diffusion coefficient we assume in our study. The plot shows the dependence on position of the coefficient for $r_T = 30$ pc and different values for $\delta$.} 
\label{Fig:transition}
\end{figure}

We assume for simplicity spherical symmetry (unlike for example what assumed in Ref.~\cite{DAngelo:2015cfw,DAngelo:2017rou} which also explored spatially dependent diffusion constants but with variations in the axial $z$ direction) and we explore two possibilities for the functional form of the radial dependence: either the transition is sudden, and modeled by the Heaviside function $\theta (r)$, or the transition is smooth; in this latter case we parametrize the radial dependence through an arctan, $\rm{tan}^{-1} (r)$ functional form. In summary, we assume the following two functional forms:
\begin{subequations}
\begin{equation}
D_{\theta}(r)=D_1 \theta(r_{T}-r) + D_2 \theta(r- r_T)
\label{eq:theta_fun}
\end{equation}
\begin{equation}
D_{\rm{T}}(r)=D_1 + \frac{(D_2 - D_1)}{\pi}\left(\rm{tan}^{-1}\left(\frac{r-r_T}{\delta} \right) + \frac{\pi}{2} \right),
\label{eq:arcT}
\end{equation}
\end{subequations}
where $r_T$ stands for the distance from the PWN at which the transition takes place, $D_1$ and $D_2$ are the diffusion constants in medium 1 and 2 respectively and $\delta$ is the width of the transition in the case of it being ``smooth''. We display our choices for the radial dependence of the diffusion coefficient in Fig. \ref{Fig:transition}. Naturally, for $\delta\to0$ the two functional forms converge -- a fact we use to cross-check our numerical results.

\section{Results: cosmic ray positrons and electrons from the Geminga pulsar}\label{sec:results}

We solve Eq.~(\ref{eq:diffeq}) using the MC method described above to compute the spectrum at Earth of cosmic rays injected by a single pulsar with a spatially dependent diffusion coefficient. For reference, we focus our study on the case of the Geminga pulsar, and employ a characteristic age of $t=3.42 \times 10^{5} $ yr, a distance to the pulsar $r=$ 250 pc, an initial injection power-law spectrum of $\alpha = 2.34$, and a total energy released  equal to $11 \times 10^{48} \rm{ergs}$, in agreement with what extrapolated by the HAWC analysis.

\begin{figure}[t]
\centering
\includegraphics[width=1\linewidth]{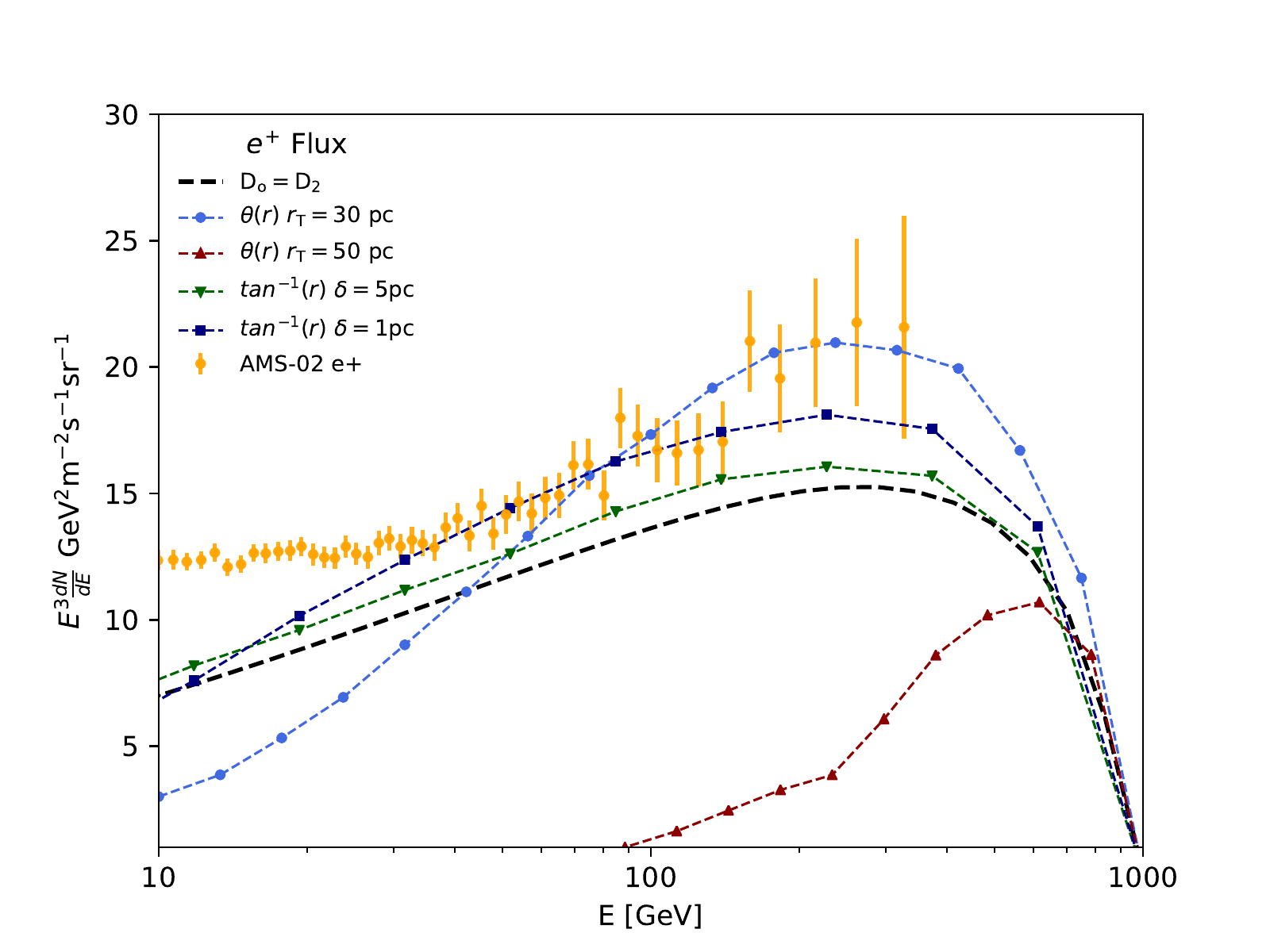}
\caption{The positron spectrum computed using the Monte Carlo algorithm described in section \ref{sec:Solution} for multiple choices for the radial dependence of the diffusion constant. We also show the AMS-02 measurement of the positron flux. We use the parameters appropriate for the Geminga pulsar, i.e. $t=342000$ yr, $r=250$ pc $\alpha=2.34$, and $D_1 = 3.86 \times 10^{26} \rm{cm}^2 \rm{s}^{-1}$ and $D_2 = 3.86 \times 10^{28} \rm{cm}^2 \rm{s}^{-1}$.}
\label{Fig:Spectrum_results}
\end{figure}
\begin{figure*}
\centering
\includegraphics[width=\textwidth]{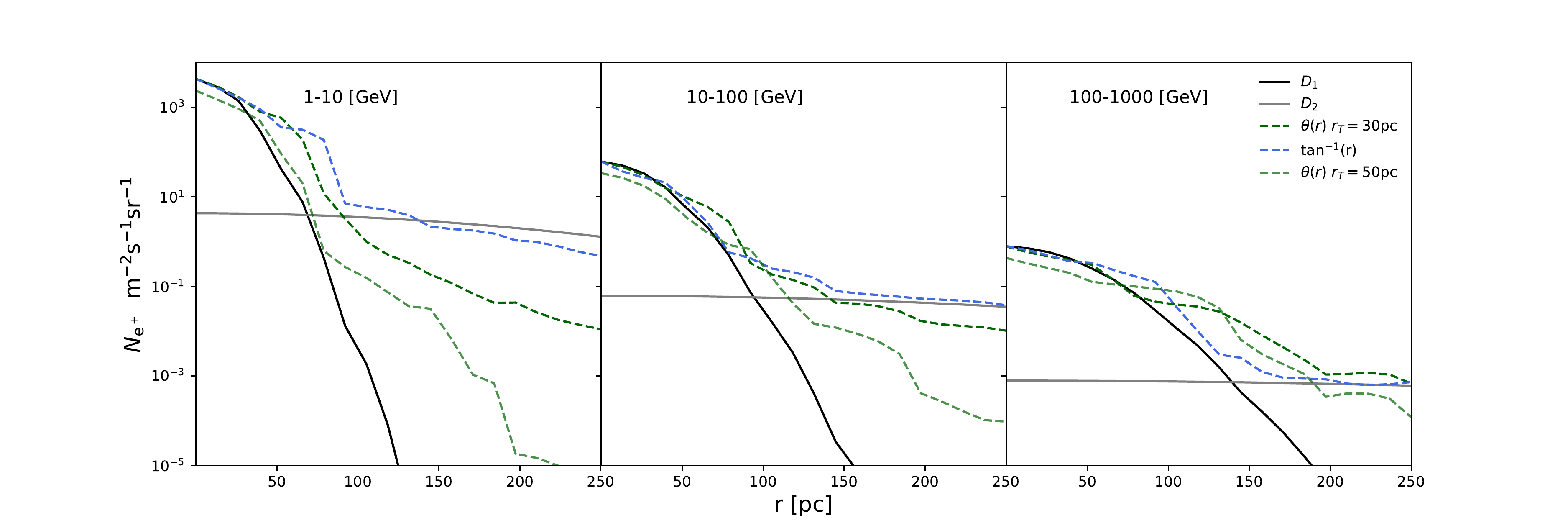}
\caption{The integrated flux per square meter per second per steradian for different energy ranges and different diffusion scenarios. The dashed lines represent the cases in which the medium undergoes a transition. $D_1 = 3.86 \times 10^{26} \rm{cm}^2 \rm{s}^{-1}$, $D_2=3.86 \times 10^{28} \rm{cm}^2 \rm{s}^{-1}$ and $t=342000$ yr.}
\label{Fig:N}
\end{figure*}
\begin{figure}
\centering
\includegraphics[width=1\linewidth]{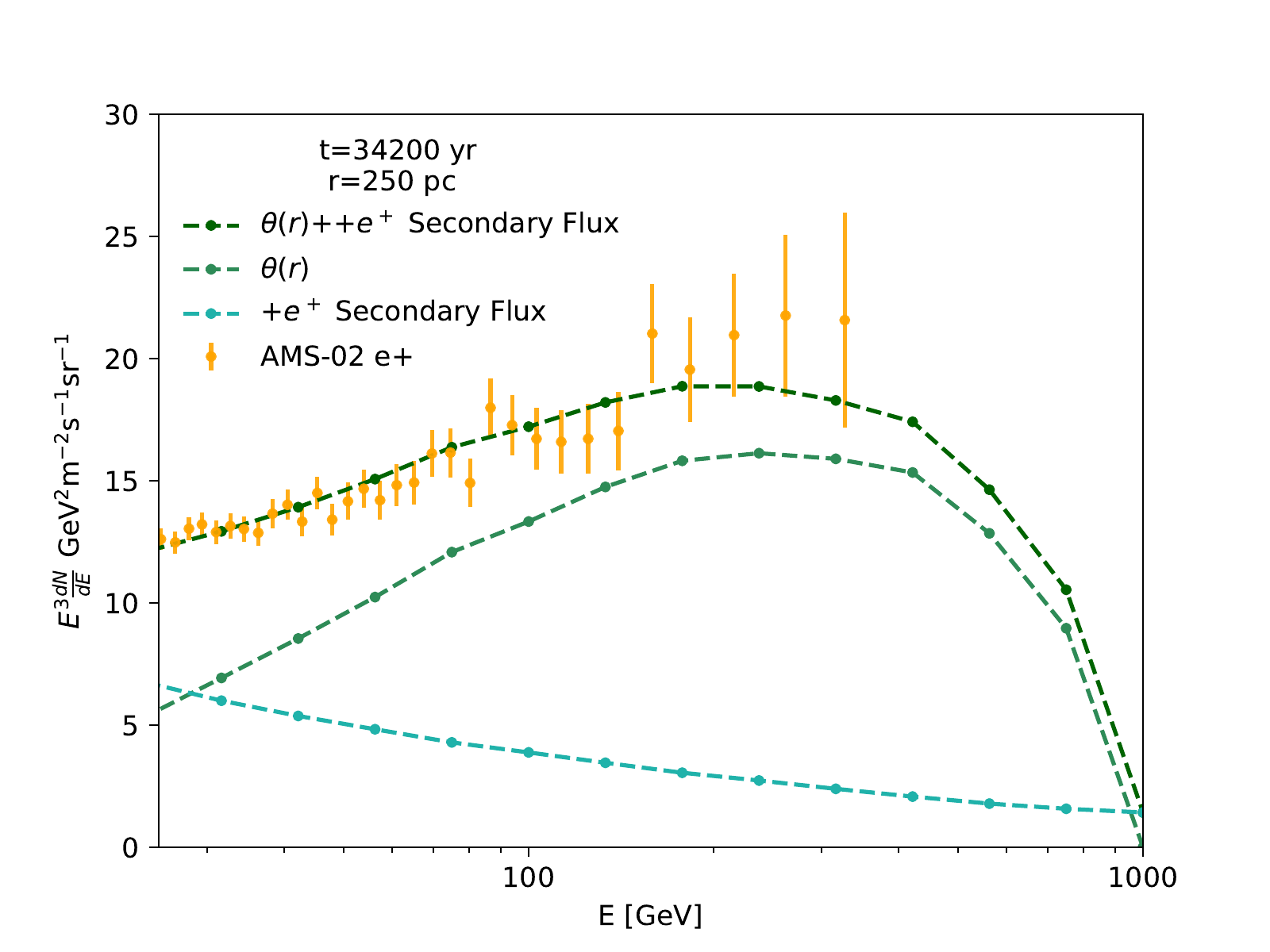}
\caption{The positron flux with the contributions of Geminga and secondary production, compared to AMS-02 data. The transition for this case is set to $35$ pc with $\sim 40\%$ $e^\pm$ energy injection efficiency. }
\label{Fig:fit}
\end{figure}

 Our main results are shown in Fig. \ref{Fig:Spectrum_results}--\ref{Fig:fit}. Fig. \ref{Fig:Spectrum_results} shows the positron flux at Earth computed for different choices for the radial dependence of the diffusion coefficient. Here, we use a diffusion coefficient inside the PWN which is a factor of 100 smaller than outside. We also show the positron flux as measured by AMS-02 \cite{PhysRevLett.110.141102}. The black dashed line indicates the case of no spatial variation in the diffusion coefficient. The blue and read lines have a step-function transition at 30 and 50 pc from the pulsar, respectively, while the green and dark blue lines show the arctan functional form with a broader and narrower transition region of 5 and 1 pc, respectively; in both cases we assume $r_T=30$ pc. 
 
The figure clearly shows how a very significant high-energy flux of positrons is to be expected from e.g. Geminga if indeed diffusion transitions from an inefficient region in the nebula to values typical of the ISM 30 to 50 pc away from the pulsar. The precise spectrum depends, as expected, on the details of the radial dependence of the diffusion coefficient, as well as on the precise location of the transition region. We find for example that a more extended inefficient diffusion region, besides the obvious overall suppression of the flux (in our case study of around a factor 2 going from 30 to 50 pc radius), also enhances the fluxes of higher-energy particles -- a consequence of the energy-dependent diffusion length we show in Fig.~\ref{Fig:diff_lenght}. A ``smooth'' transition tends to populate lower-energy particle fluxes versus a sharper transition, as visible comparing the 5 and 1 pc widths of the arctan cases, and also the arctan cases versus the sharp transitions.

An additional notable feature of fig.~\ref{Fig:Spectrum_results} is that for the same total energy injected in electrons and positrons, generally it is not true that a region of inefficient diffusion around the pulsar suppresses the flux at Earth. Rather, we find that for several choices for the radial dependence of the diffusion coefficient, the positron flux from Geminga is actually {\it enhanced} compared to a uniform, large diffusion coefficient. 

We explore and explain this feature by studying, in fig.~\ref{Fig:N}, the integrated flux for different energy ranges as a function of distance. We can appreciate from these plots how the positron number density between 100 GeV and 1000 GeV is higher than the constant diffusion case, as we find in the spectral plot. Our interpretation of these results is that the positrons in a high diffusion constant region reach further distances than the case where a transition is present: Positrons that were injected in a low diffusion constant region are effectively ``trapped'' until later times and thus, when they reach the high diffusion region, they do not have enough time to effectively diffuse. The greater the distance at which the transition happens, the more the spectrum resembles that of the low diffusion constant.

 Fig.~\ref{Fig:fit} provides evidence that a pulsar interpretation of the anomalous positron fraction is perfectly consistent with inefficient diffusion around PWNe. The figure utilizes a transition region with a steep transition profile and of transition radius 35 pc, and an assumed fraction of the pulsar spin-down luminosity converted to electron-positron pairs consistent with the power inferred by the TeV observations of HAWC, i.e. around $40\%$ \cite{Abeysekara:2017old}. We also show a standard secondary positron flux (light blue), and the sum of the positrons from secondary processes plus those from Geminga. The figure clearly shows that with a radially dependent diffusion coefficient positrons from Geminga can dominate the positron flux measure at Earth and give an acceptable fit to the data (notice that for simplicity and for the sake of clarity we do not include additional pulsars, which certainly contribute to the high-energy positron flux as well, see e.g. \cite{Hooper:2017gtd}).
 
 \section{Macroscopic effects of inefficient diffusion in PWNe} \label{sec:Macroscopic_effects}

In this section we quantify the degree to which the implied inefficient diffusion coefficient inside PWNe, $D_{\rm PWN}\sim 10^{-2}D_{\rm ISM}$ affects cosmic ray diffusion in the Galaxy at macroscopic scales. While we postpone detailed simulations and observational tests for future work, here we employ observationally-motivated relations between the PWN radius, $R_{\rm PWN}$, and the pulsar characteristic age, ${\tau}_c$ from the Australia Telescope National Facility (ATNF) catalogue of Galactic pulsars\footnote{http://www.atnf.csiro.au/people/pulsar/psrcat/}, to infer the total volume in the Galactic diffusive halo where inefficient diffusion is expected, $V_{\rm PWN}$.

PWNe are generated within the supernova remnants accompanying the birth of pulsars, and here we intend to estimate their radial size evolution as a function of time. We consider three distinct phases of the evolutionary stages of the PWNe following the procedure of Ref.~\cite{Abdalla:2017vci}. Initially, we assume constant energy output $\dot{E} \sim \dot{E_0}$ while $t < {\tau}_0$, where $t$ is the age of the pulsar and ${\tau}_0$ is the initial spin-down time scale of the pulsar. During this stage, the pulsar is surrounded by the un-shocked ejecta of the surrounding SNR and the PWN expands at supersonic speeds.

As the surrounding SNR accumulates matter from the ISM, the SNR ejecta are decelerated,  triggering an inward moving reverse shock in the SNR which will then collide with the expanding PWN, causing the expansion of the PWN to slow down. This stage occurs at some reverse-shock interaction time, $t_{\rm rs}$.

After this dynamical period of collision between the PWN and SNR reverse-shock, the PWN again steadily expands, making the assumption $t_{rs} > {\tau}_0$; at this stage it is no longer warranted to take $\dot{E} \sim \dot{E_0}$, and thus the expansion slows down further. It should be noted that this is a simplified model of PWNe radial evolution; a more detailed picture of the evolution can be found in \cite{Gaensler:2006ua}. Finally, we  use here the following equations to estimate $R_{\rm PWN}$ as a function of time:

\[
    R_{\rm PWN}(t)\propto\left\{
                \begin{array}{ll}
                  t^{6/5}, \text{for $t \leq {\tau}_0$} \\
                  t, \text{for ${\tau}_0 < t \leq t_{rs}$}\\
                  t^{3/10}, \text{for $t > t_{rs}$},
                \end{array}
              \right.
\]
where $t$, ${\tau}_0$, and ${\tau}_c$ are related via the following equation \cite{Gaensler:2006ua}:
\begin{equation}
{\tau}_c \equiv \frac{P}{2\dot{P}} = ({\tau}_0 + t)\frac{n-1}{2},
\end{equation}
$P$ being the spin period, $n$ is the braking index of the  pulsar, ${\tau}_0$, $t_{\rm rs}$, and these values are listed in \cite{Abdalla:2017vci}. \\

\begin{figure*}
\centering
\includegraphics[width=0.8\textwidth]{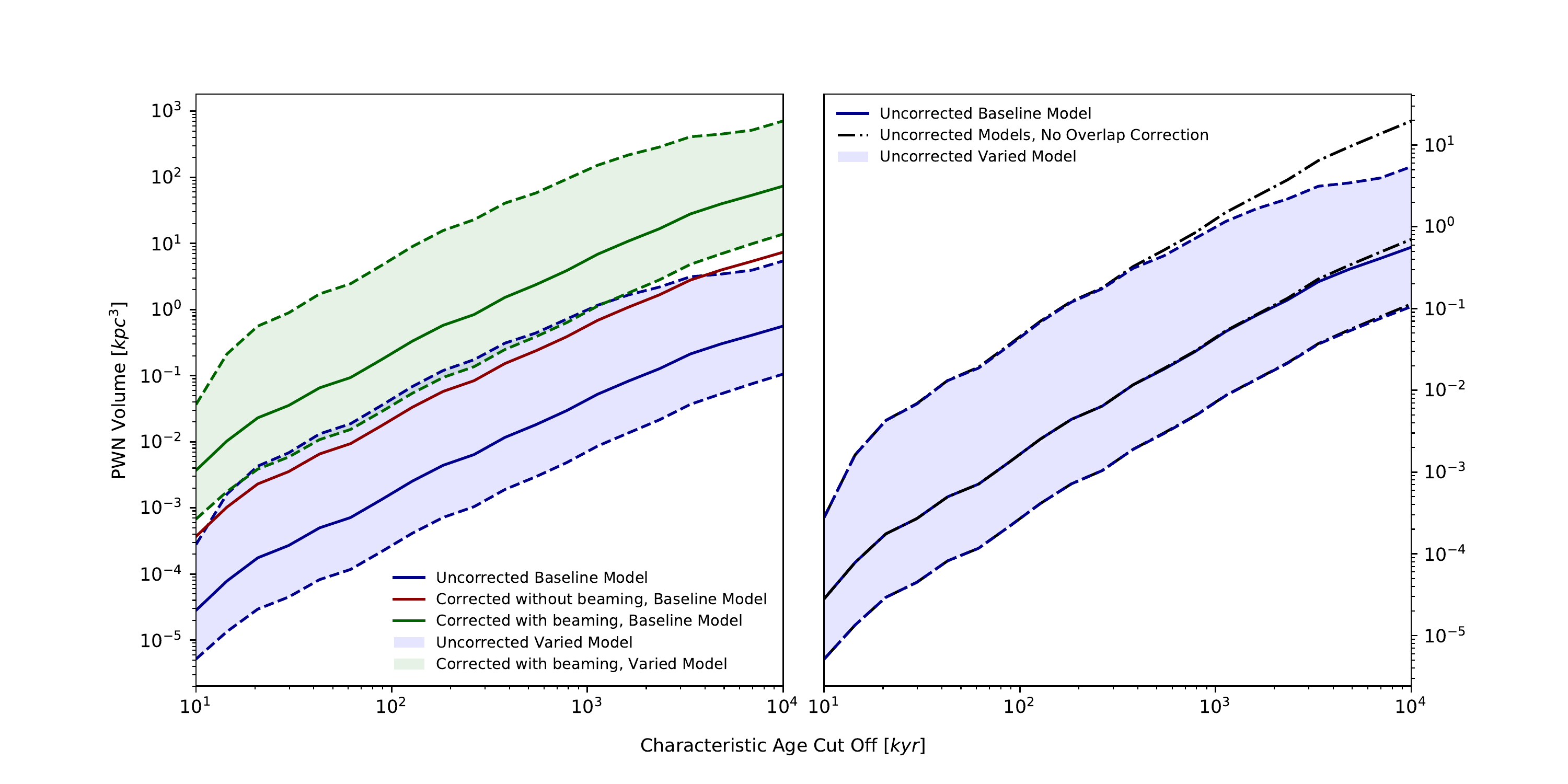}
\caption{We plot the net PWNe volume versus the pulsar characteristic age cut off with constants given by the baseline and varied models for PWNe radius evolution as listed in \cite{Abdalla:2017vci}. On the left, the blue region gives the total volume of the PWNe of pulsars sampled from the ATNF catalogue, the red curve is the baseline model adjusted for incompleteness factors in pulsar sampling that account for distance dependent and galactic longitude dependent selection effects\cite{Yusifov:2004fr}, and the green region accounts for selection effects due to a pulsar mean beaming fraction of $\langle f\rangle \sim 0.10$\cite{1998MNRAS.298..625T} in addition to the same selection effects as the red curve. On the right, the incomplete volume is plotted with a curve that does not correct for overlapping PWNe to depict the amount of overlap of PWNe at different characteristic age cut offs.}
\label{Fig:vol_vs_age}
\end{figure*}

The volume of each putative PWN was calculated and then summed to derive the total volume of PWNe in the Milky Way, $V_{\rm PWN}$. We note that it was necessary to account for overlapping PWNe, and so when the volumes were summed, each contiguous region of connected PWNe was checked. If a PWN was isolated, the volume could be calculated in the simple case just using its radius. If two PWNe were connected, their volumes were summed analytically and the overlap region was subtracted to avoid double counting. To account for regions with 3 or more overlapping PWNe, a mesh subdivided into small cubes was created for each contiguous region. We checked these cubes to mark which fall within the radii of one of the PWNe, and the total volume was then estimated by summing these.  The net PWNe volume as a function of characteristic age cut off is plotted in Fig. \ref{Fig:vol_vs_age}.

At the time of our sampling from the ATNF catalogue, 69.4\% of pulsars had the necessary published values to estimate the volumes of their PWNe, leaving us a total sample of 1830 pulsars. To account for the incompleteness, we referred to previous works where a distribution of pulsars as a function of the center of the Galaxy has been suggested \cite{Yusifov:2004fr}; to estimate the total number of pulsars in the Galaxy, we then integrated their radial distribution:
\begin{equation}
\rho(R) = A\left(\frac{R+R_1}{R_{\odot} + R_1}\right)^aexp\left[-b\left(\frac{R - R_{\odot}}{R_{\odot} + R_1}\right)\right],
\end{equation}
where ${\rho}(R)$ is a cylindrical surface density with constants $A = 37.6 \pm 1.9$ $ {\rm kpc}^{-2}$, $a = 1.64 \pm 0.11$, $b = 4.01 \pm 0.14$, $R_1 = 0.55 \pm 0.10 $ kpc, and $R_{\odot} = 8.5$ kpc, where $R_{\odot}$ is the Sun-Galactic Center (GC) distance. Ref. \cite{Yusifov:2004fr} derives the functional form of this distribution by sampling the radial distribution of pulsars in the ATNF catalogue, excluding binary, recycled ($\dot{P} < 10^{-17}$), globular cluster and Large and Small Magellanic Cloud pulsars. At the time, this included 1254 pulsars in total.  Correction factors to relate this observed distribution to the total distribution of pulsars in the Milky Way accounted for direction-dependent $K(l)$ and distance-dependent $K(r)$ selection effects:
\begin{equation}
{\rho}(r,R,l(r,R)) = K(l)K(r){\rho}_o(r,R,l(r,R)),
\end{equation}
${\rho}_o$ being the observed surface density of pulsars, $r$ and $R$ the heliocentric and galactocentric radii (respectively), and $l$ being the Galactic longitude. Direction-dependent selection effects are accounted for by estimating the effect of background radiation on survey sensitivity as a function of $l$. Distance-dependent selection effects account for minimum detectable flux and pulse broadening, scattering and scintillation. Integrating (12) gives a total pulsar number of $(24 \pm 3) \times 10^3$. Accounting for a mean beaming fraction of $\langle f\rangle \sim 0.10$   \cite{1998MNRAS.298..625T}, the total pulsar number is $(240 \pm 30) \times 10^3$. Fig.\ref{Fig:vol_vs_age} shows the uncorrected baseline model with standard deviation as well as the model of the corrected volume (with and without accounting for beam fraction).

Simpler estimates of the volume fraction of inefficient diffusive regions have been made in the past. For example, Hooper and Linden \cite{Hooper:2017gtd} used one putative single value for the typical size of PWNe of 30 pc, and a typical supernova rate for the Galaxy (0.03 per year) to infer that, for a 20 kpc radius and 200 pc half-height for the diffusion region, the total fraction of inefficient diffusion volume is of about 0.007, or about 3.5 kpc$^3$. This estimate is entirely consistent with our estimates, which indicate similar volumes for the baseline model corrected with beaming effects, for a characteristic age up to roughly 1 Myr.

What is the impact of pockets of inefficient diffusion in the Galaxy, and which diagnostics can be used to detect them? Ref.~\cite{DAngelo:2017rou} addressed this question by studying the diffuse gamma-ray emission resulting from cosmic rays trapped for longer-than-usual times inside the ``near-source region'' where they were accelerated; inelastic collisions of cosmic-ray protons with nuclei in the interstellar medium leads via neutral pion decay to a possibly significant gamma-ray emission from these ``halos'' with large self-contained cosmic-ray populations. Ref.~\cite{DAngelo:2017rou} then proceeds to studying the integrated diffuse gamma-ray emission in various angular regions. The key issue with this diagnostic is that the emission is basically indistinguishable, spectrally or morphologically, from other sources of diffuse emission; for example, in Ref.~\cite{Carlson:2016iis,Carlson:2016iis} it was shown how models of the diffuse emission with an enhanced cosmic-ray source population tracing regions of hypothetically large star formation provide a better global fit to the observed Galactic diffuse gamma-ray emission; such a contribution would be likely highly degenerate with the extended halos considered in Ref.~\cite{DAngelo:2017rou}

Before we turn to the observational tools we propose to test whether or not cosmic rays undergo inefficient diffusion near their acceleration sites, we shall discuss first the typical confinement times in these regions (notice that Ref.~\cite{DAngelo:2017rou} also discusses escape times, albeit in a completely different diffusion scheme, see their figure 2). We notice that since the mean free path for a particle in the diffusive regime is of the order
$$
\langle L \rangle \sim \sqrt{D\cdot t},
$$
the ratio of the residence time for cosmic rays in the diffusive regime inside inefficient diffusion regions to that inside the rest of the ISM, is of the order
\begin{equation}
\frac{t_{\rm PWN}}{t_{\rm ISM}}\sim\left(\frac{\langle V \rangle_{\rm PWN}}{\langle V \rangle_{\rm ISM}}\right)^{2/3}\ \frac{D_{\rm ISM}}{D_{\rm PWN}}\sim 10^{2}\ \left(\frac{\langle V \rangle_{\rm PWN}}{\langle V \rangle_{\rm ISM}}\right)^{2/3}
\end{equation}
Thus, for effective diffusive volumes up to 1,000 times larger than our estimated inefficient diffusion volume, cosmic rays in the diffusive regime would spend {\it more} time inside regions of inefficient diffusion than elsewhere. For a diffusive halo of total volume
\begin{equation}
\langle V \rangle_{\rm ISM}\simeq 2500\ {\rm kpc}^3\left(\frac{R_h}{20\ {\rm kpc}}\right)^2\left(\frac{z_h}{1\ {\rm kpc}}\right)
\end{equation}
this means that for $\langle V \rangle_{\rm PWN}\gtrsim2.5\ {\rm kpc}^3$ cosmic rays are likely to spend more time trapped within acceleration sites than in the ISM at large. From our fig.~\ref{Fig:vol_vs_age} this, in turn, corresponds to very likely PWN sizes, corresponding to pulsar ages (for our baseline model corrected for incompleteness, and with beaming corrections) near approximately few $\times10^6$ yr of age, thus close to the age of the observed TeV halos of Geminga and Monogem \cite{Abeysekara:2017old}.

We note that the size of the diffusive halo height as determined from studies of Galactic magnetic fields \cite{Beck:1995zs}, from diffuse gamma-ray observations \cite{Ackermann:2012pya}, radio observations \cite{Bringmann:2011py, DiBernardo:2012zu, Orlando:2013ysa}, and secondary-to-primary ratios \cite{Reinert:2017aga} is typically larger than the reference value we employ above ($z_h\sim 1\ {\rm kpc}$), and is in fact larger than at least 4 kpc \cite{Reinert:2017aga} with typical values around 10 kpc \cite{Ackermann:2012pya, Orlando:2013ysa}. However, such large values are only relevant for cosmic-ray species which loose energy inefficiently, such as hadronic cosmic rays, and which thus indeed sample large swaths of the diffusive halo during their residence time. For cosmic-ray electrons, instead, as illustrated in our fig.~\ref{Fig:diff_lenght}, the effective diffusion length is always below 1 kpc, even for very large diffusion coefficient choices (notice that this statement depends on the age of the pulsar under consideration, according to Eq.~(\ref{eq:rdiffmax}); for the largest diffusion coefficient we consider here, the maximal diffusion radius corresponds to 0.8 kpc for pulsars of age $t_o=10^6$ yr and to 1.7 kpc for pulsars of age $t_o=10^7$ yr, the largest characteristic age we consider in fig.~\ref{Fig:vol_vs_age}); in turn, this implies that cosmic-ray electrons do not travel further than around $1-2$ kpc from their acceleration sites, which lie dominantly on the Galactic disk, at $z\sim0$. The effective diffusive halo for cosmic-ray electrons is thus at most of height $z_h\lesssim 1-2$ kpc. As indicated, for hadronic cosmic rays this does not hold; however, any observational diagnostic  relevant here (to be discussed below) depends on inelastic processes that require large gas densities; such regions (those with enough target gas density) lie also at very low Galactic latitudes, hence, again, the effective diffusive halo is much smaller than what would implied by $z_h\sim 10$ kpc.

How can this prediction be tested? The key observables are related to indirect probes of high-energy cosmic rays, such as for example synchrotron and inverse Compton energy losses of high-energy electrons and positrons, or inelastic hadronic processes producing neutral pions. If high-energy cosmic rays indeed spend a significant amount of time in relatively small pockets of inefficient diffusion, they will proportionally lose appreciably more energy in those environments than in the standard case with homogeneous diffusion. As a result, an appropriate diagnostic would consist of a study of the clustering scales of emission at the relevant wavelengths (ranging from radio for synchrotron emission, to X-ray, to soft and hard gamma rays for inverse Compton). The relevant angular scales are of the order of
\begin{equation}
\theta\sim\frac{R_{\rm PWN}}{d_{\rm PWN}},
\end{equation}
and thus range from around 5 degrees in the case of the PWN observed by HAWC, down to ${\cal O}\left(0.5^\circ\right)$ for pulsars near the center of the Galaxy.  

The prediction for this model thus consists of significantly greater emission power on physical scales of the order of PWN sizes, and on angular scales extending from 0.1 to several degrees, as well as a brighter emission in the Galactic regions where more abundant populations of PWNe are present. The emission is expected over a broad range of frequencies, from very high energy gamma rays, down to soft gamma rays, X-rays, and radio frequencies. The diagnostic will consist of, for instance, calculating the angular power spectrum predicted for significant emission from PWN (utilizing both the ATNF pulsar catalogue and corrections thereof) and following the procedure outlines in Ref.~\cite{Fornasa:2016ohl} to constrain dark matter emission. A similar procedure was used to describe how to constrain decaying dark matter emission at X-ray frequencies with eROSITA in Ref.~\cite{Zandanel:2015xca}

In addition to the angular power spectrum, other possible diagnostics include (i) a wavelet analysis, as pioneered again for the case of the diffuse gamma-ray emission in Ref.~\cite{Bartels:2015aea} and there used to distinguish the clustering properties of millisecond pulsars versus dark matter, or (ii) using metrics like the flux probability density function which Ref.~\cite{Lee:2014mza} utilized to distinguish uniform from non-uniform diffuse emission in the gamma-ray sky. At present, all of the outlined methods are under investigation in their application to the problem of disentangling a contribution from in source, ``trapped'' cosmic-ray emission from the diffuse emission.


\section{Discussion}
\label{sec:Discussion}
In this work we have explored the possibility of a non-homogeneous medium through which cosmic rays diffuse in the vicinity of their acceleration sites as a solution to the low diffusion coefficient inferred by the HAWC Collaboration \cite{Abeysekara:2017old} compared to what inferred from global fits to Galactic cosmic-ray data \cite{Strong:1998pw}. We have implemented a numerical solution to the problem of spherically-symmetric diffusion with a radially-dependent diffusion coefficient, which consists of a Monte Carlo integration. We validated our numerical procedure against known analytical solutions in the appropriate limits, and we then proceeded with the exploration of a variety of functional forms for the transition from low to high efficiency in the cosmic-ray diffusion away from their injection sources.

We found that (i) the sharper the transition from inefficient to efficient diffusion, the more abundant the asymptotic relic population of high-energy cosmic rays; (ii) for the specific case of Geminga, the flux of cosmic rays predicted by scenarios of inefficient-to-efficient diffusion versus uniformly efficient diffusion, with the same injected flux, can actually be larger, especially at higher energies; (iii) the location of the transition region from inefficient to efficient diffusion significantly affects the asymptotic cosmic-ray spectrum, with a larger suppression of low-energy cosmic rays for larger transition radii; (iv) there exist several realization of radially-dependent diffusion from Geminga or Monogem that suitably explain the excess high-energy positron flux at Earth as originating from those nearby, middle-aged pulsars.

We then proceeded to estimate and discuss the {\it global} implications for Galactic cosmic-ray transport of the existence of pockets of inefficient diffusion around cosmic-ray accelerator. We showed a data-driven approach, based on the ATNF pulsar catalogue and on completeness corrections thereof, and on observationally and theoretically motivated models for the physical size of pulsar wind nebulae, to estimate the total Galactic volume with inefficient diffusion. We showed that such volume is likely at least 1\% of the Galactic diffusive halos, and possibly larger. If this is the case, we argued that cosmic rays spend most of their residence time inside inefficient diffusion regions.

We proposed several observational diagnostics to validate our findings, which essentially plan to utilize a study of diffuse emission power on the angular scales relevant for the expected size of regions of inefficient diffusion, ranging from a fraction of a degree to several degrees in the sky.

In conclusion, the key lesson from the HAWC results \cite{Abeysekara:2017old} is evidence that diffusion is not homogeneous in the Galaxy, and that cosmic-ray diffusion is inefficient inside pulsar wind nebulae; this is not an unexpected result, and it had been anticipated by several theoretical studies on self-generated turbulence inside cosmic-ray acceleration sites \cite{Malkov:2012qd}. Unlike what claimed in Ref.~\cite{Abeysekara:2017old}, we showed that for several choices for the transition from inefficient to efficient diffusion, nearby pulsars such as Geminga are likely very significant contributors to the local high-energy positron flux. An important implication of the HAWC results is that likely cosmic rays spend most of their Galactic residence time inside hypothetical inefficient diffusion regions. We outlined observational methods to test this possibility which include studying the angular power spectrum of diffuse emission at a variety of wavelengths, wavelet analyses, and non-Poissonian photon statistics.\\

{\it Note Added.} During the final stages of this work, Ref.~\cite{Fang:2018qco} appeared, which deals with a toy model of a two-zone diffusion coefficient and a semi-analytic solution consisting of simple matching of boundary conditions; the results of Ref.~\cite{Fang:2018qco} agree with ours when directly comparable, which is limited to the case when the diffusion coefficient has a step-like discontinuity.\\

\appendix
\section{Results of the Monte Carlo Simulation}

\begin{figure}[ht!]
\centering
\includegraphics[width=1\linewidth]{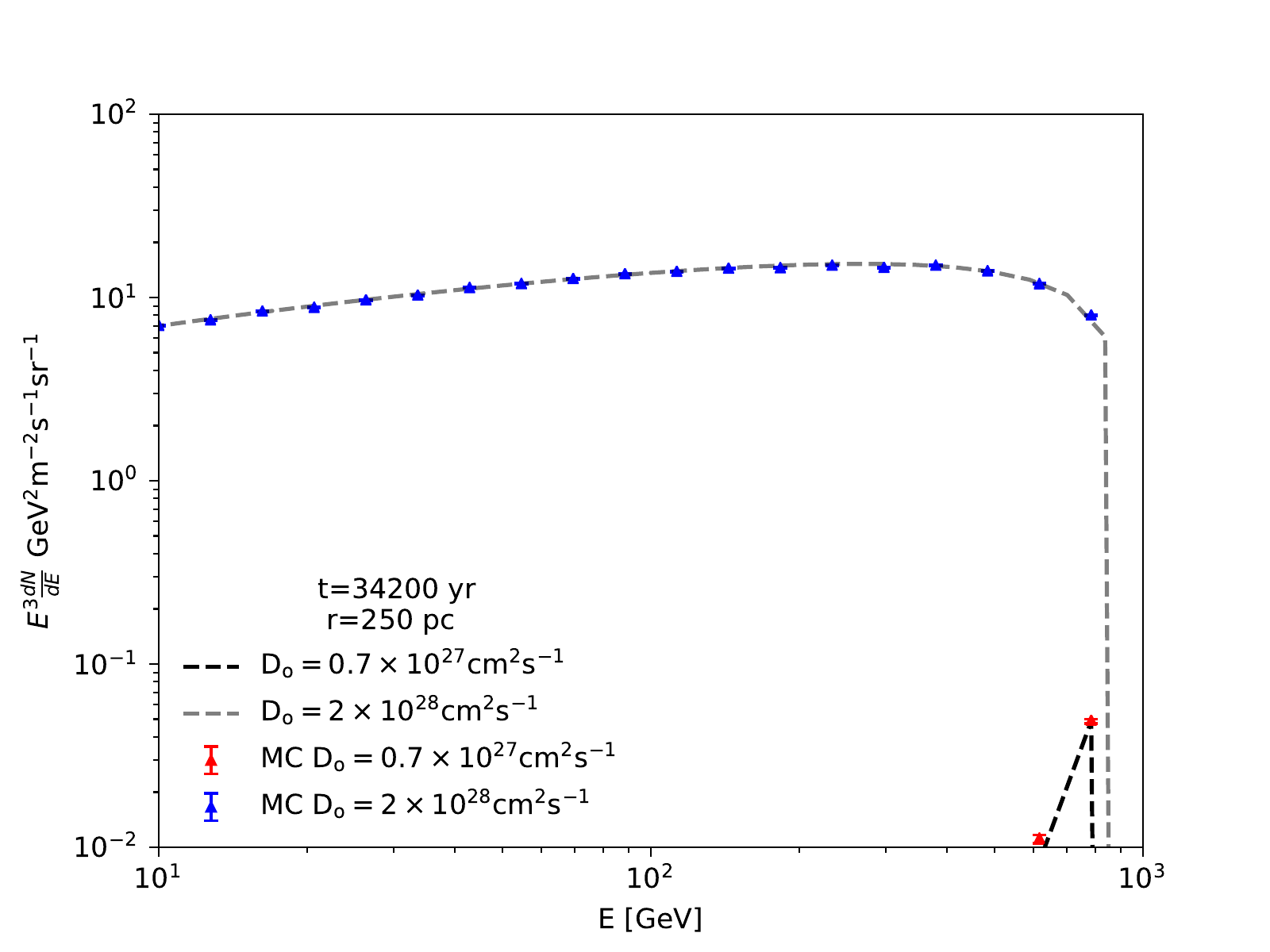}
\caption{The accuracy of the Monte Carlo Method. The dashed lines represent the analytical solution of the diffusion equation in the case of a $\delta$-like initial injection, the blue and red points were generated through the MC simulation as described in section \ref{sec:Solution}. The error bars displayed the probability of finding the point $1 \sigma$ away from the true value. The results indicate that our algorithm is in good agreement with the analytical solutions.} 
\label{Fig:MC_method}
\end{figure}

As described in Section \ref{sec:Solution}, we used a Monte Carlo (MC) algorithm to solve the 3D diffusion equation by using a Box-Muller transformation. In order to test our method we solved the equation and compared it with the analytical very well known solution Eq. \ref{eq:analytical_solution}. Since the MC method is purely stochastic, we computed the $1 \sigma$ error for the points generated. The results are presented in Fig. \ref{Fig:MC_method} for the values of the diffusion regime used through this work. In the figure, the dashed lines indicate the analytical solution for a delta-function injection source, while the blue and red data points show the results of our MC method. The numerical accuracy of our algorithm is clearly very high.\\


\section*{Acknowledgements}
We would like to thank Nicolas Fernandez for helpful discussions. This work was funded by a UCMEXUS-CONACYT collaborative project. SP is partly supported by the U.S. Department of
Energy grant number de-sc0010107. JR acknowledges financial support from CONACYT and CONACYT project 182445.

\bibliography{bib.bib}
\end{document}